\documentclass{emulateapj}
\usepackage{graphicx}

\newcommand{\unit}[1]{\ifmmode \:\mbox{\rm #1}\else \mbox{#1}\fi}

\newcommand{\mone}{\mm{^{-1}}}

\newcommand{\kms}{\unit{km~s\mone}}

\newcommand{\mm}[1]{\mbox{$#1$}}

\newcommand{\egal}{ESO325--G004}

\begin{document}
\title{Discovery of Strong Lensing by an Elliptical Galaxy at z=0.0345\footnotemark[1]}\footnotetext[1]{Based on observations made with the NASA/ESA Hubble Space Telescope, obtained at the Space Telescope Science Institute, which is operated by the Association of Universities for Research in Astronomy, Inc., under NASA contract NAS 5-26555. These observations are associated with program 10429.}
\author{Russell J. Smith\altaffilmark{2}}
\altaffiltext{2}{Department of Physics, University of Waterloo, 200 University Avenue West, Waterloo, Ontario N2L 3G1, Canada. rjsmith@astro.uwaterloo.ca}
\author{John P. Blakeslee\altaffilmark{3}}
\altaffiltext{3}{Department of Physics and Astronomy, The Johns Hopkins University,  3400 North Charles Street, Baltimore, MD 21218, USA. jpb@pha.jhu.edu}
\author{John R. Lucey\altaffilmark{4}}
\altaffiltext{4}{Department of Physics, University of Durham, Durham DH1 3LE, United Kingdom. john.lucey@durham.ac.uk}
\author{John Tonry\altaffilmark{5}}
\altaffiltext{5}{Institute for Astronomy, University of Hawaii, 2680 Woodlawn Drive, Honolulu, HI 96822-1897, USA. jt@ifa.hawaii.edu}

\slugcomment{Accepted for publication in Astrophysical Journal Letters}
\shortauthors{Russell J. Smith et al.}
\shorttitle{Lensing by an elliptical galaxy at z=0.0345}

\begin{abstract}
We have discovered strong gravitational lensing by the galaxy \egal, in images obtained with the {\it Advanced Camera for Surveys} on the {\it Hubble Space Telescope}. The lens galaxy is a boxy group-dominant elliptical at $z=0.0345$, making this the closest known galaxy-scale lensing system. The lensed object is very blue ($B-I_c\approx1.1$), and forms two prominent arcs and a less extended third image. The Einstein radius is  $R_{\rm Ein}=1.9$\,kpc ($\sim3$\,arcsec on the sky, cf. 12\,arcsec effective radius of the lens galaxy). Assuming a high redshift for the source, the mass within $R_{\rm Ein}$ is $1.4\times10^{11}\,M_\odot$, and the mass-to-light ratio is $1.8 (M/L)_{\odot,I}$. The equivalent velocity dispersion is $\sigma_{\rm lens}=310$\,\kms, in excellent agreement with the measured stellar dispersion $\sigma_v=320$\,\kms. Modeling the lensing potential with a singular isothermal ellipse (SIE), we find close agreement with the light distribution. The best fit SIE model reproduces the ellipticity of the lens galaxy to $\sim$10\%, and its position angle within 1$^\circ$. The model predicts the broad features of the arc geometry as observed; the unlensed magnitude of the source is estimated at $I_c\sim23.75$. We suggest that one in $\sim$200 similarly-massive galaxies within $z<0.1$ will exhibit such a luminous multiply-imaged source. 
\end{abstract}

\section{Introduction}

Gravitational lensing of background sources can yield valuable information on the mass profiles of galaxies, groups and clusters. In contrast to other methods, lensing constraints are independent of assumptions about the dynamical or hydrodynamic state of tracer material (e.g. stars, galaxies, X-ray gas). On galaxy scales, lensing helps to lift the degeneracy between the potential and the orbital anisotropy which plagues dynamical mass estimates. For distant galaxies, lensing constrains the mass enclosed at large radii, beyond the reach of stellar dynamical studies (e.g. Treu \& Koopmans, 2004).

The optimal configuration for lensing is with the deflecting potential at half the distance to the source.
Hence lensing is usually observed for systems at intermediate redshift, $z\sim0.3$, where the number of potential background sources is very large. At low redshifts ($z\la0.1$), strong lensing systems are rare.
On cluster scales, a number of lensed arcs have been discussed
(e.g. Blakeslee et al. 2001).
The nearest known galaxy-scale strong-lensing system, prior to this {\it Letter},
was Q2237+0305 (Huchra et al. 1985), a four-image QSO lensed by a $z=0.039$ spiral. Among galaxy lenses with extended arcs, the lowest known lens redshift is $z=0.205$, for SDSS1402+6321 (Bolton et al. 2005). This system was discovered on the basis of anomalous emission lines in the lens galaxy spectrum. A more direct approach to finding extended arcs is, of course, through imaging observations. Blakeslee et al. (2004) have discussed the excellent prospects for serendipitous discovery of strong lens galaxies with the {\it Advanced Camera for Surveys} (ACS) on the {\it Hubble Space Telescope}.

In this {\it Letter}, we report ACS discovery of a new galaxy-scale lens, with multiple extended images, at $z=0.0345$. To our knowledge, the elliptical \egal\ is the nearest known strong lensing galaxy, and provides for the first time a low-redshift analogue of distant galaxy-scale arc systems. 

We adopt the WMAP cosmological parameters, i.e. $H_0=71$\,\kms\,Mpc$^{-1}$, $\Omega_{\rm m}$=0.27, $\Omega_{\Lambda}$=0.73 (Bennett et al. 2003).

\section{The lens galaxy ESO325--G004}\label{sec:envir}

The massive boxy elliptical galaxy \egal\ (13$^h$43$^m$33$\fs$20, $-$38$^\circ$10$^m$33$\farcs$6) lies at the center of the poor cluster Abell S0740 (Abell, Corwin \& Olowin 1989), and is probably the dominant galaxy of that group. The galaxy has radial velocity $cz=$10420\,\kms\ in the Cosmic Microwave Background frame (Smith et al. 2000), corresponding to an angular scale of 0.678\,kpc\,arcsec$^{-1}$.
The galactic extinction is $E(B-V)=0.06$ (Schlegel, Finkbeiner \& Davis 1998).
The stellar velocity dispersion of \egal\ was measured by Smith et al. (2000), within an aperture 3.8$\times$3.0\,arcsec$^2$; the average of their two measurements is $\sigma_v=320\pm7$\,\kms. Smith et al. (2001) report the effective radius as $R_{\rm eff}=12.5$\,arcsec (8.5\,kpc) from R-band imaging.

\begin{figure}
\includegraphics[width=85mm]{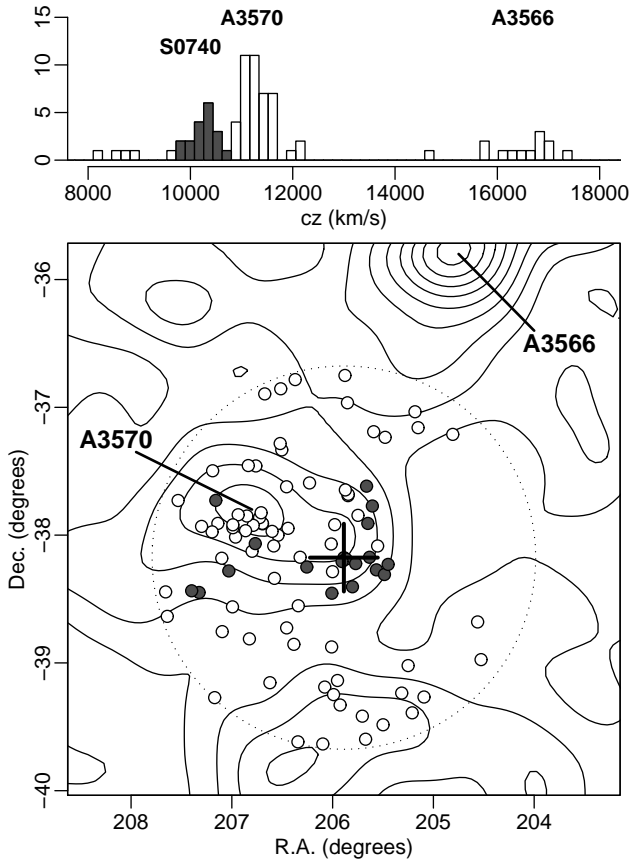}
\caption{Environment of the lensing galaxy. The main figure shows contours of projected galaxy density from the 2MASS extended source catalog. Small points mark galaxies with measured redshifts in the NASA Extragalactic Database, within 1.5$^\circ$ of \egal. Galaxies with redshifts within 500\,\kms\ of \egal\ are shaded.
The histogram above shows the same redshifts, with the same redshift interval highlighted.
}
\label{fig:envir}
\end{figure}

Understanding the environment of the lens galaxy can be critical for correct interpretation of lensing constraints. Figure~\ref{fig:envir} shows the projected galaxy density around \egal, and the distribution of published redshifts. Although not complete, the redshift data suggest that S0740 is distinct from the neighboring cluster Abell 3570 ($\sim$40\,arcmin away, and with mean redshift $\sim$1000\,\kms\ larger).

\begin{figure}
\includegraphics[width=85mm]{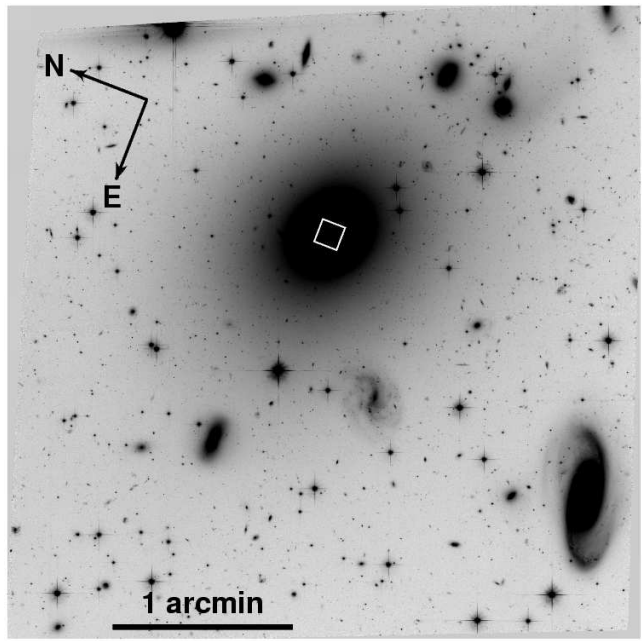}
\caption{The deep (18882\,s) F814W image of \egal\ and the surrounding field. The white square indicates
the central 8$\times$8\,arcsec$^2$ region covered by Figure~\ref{fig:lensmod}.}
\label{fig:image}
\end{figure}

\begin{figure}
\includegraphics[width=77mm]{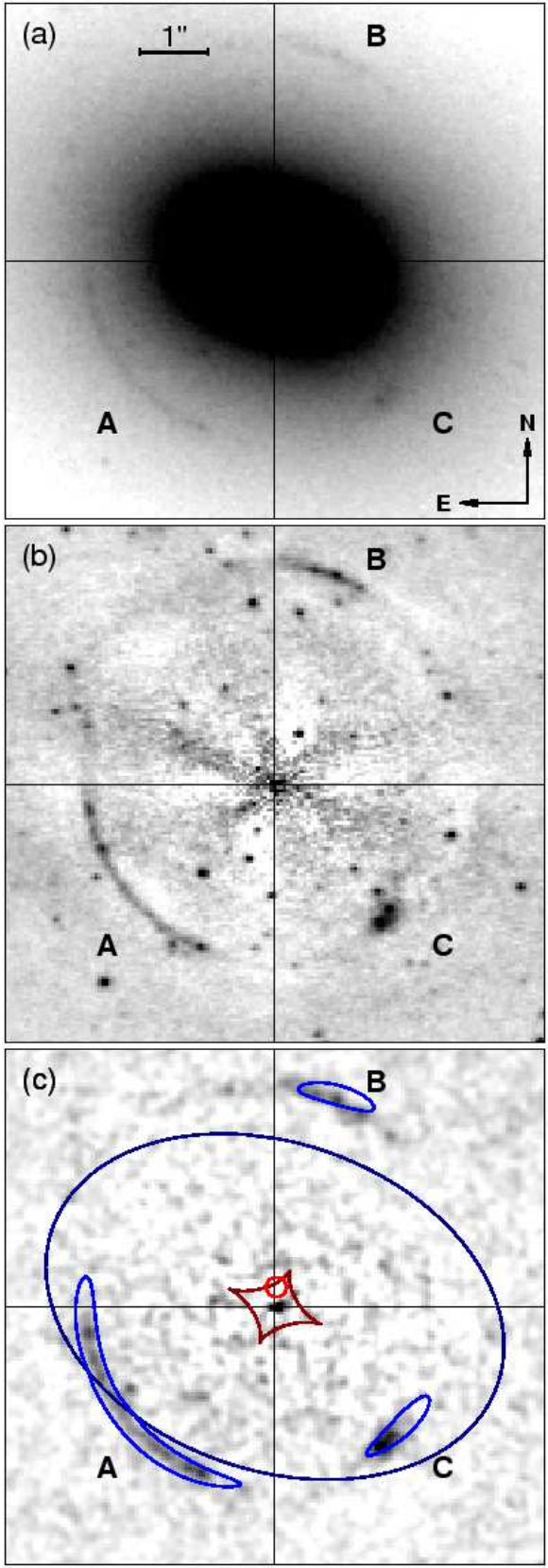}
\caption{Observations and models of lensing in \egal. Panel (a) shows the 1101\,s F475W image, with color map optimized to show the arcs. In Panel (b), the 18882\,s F814W image is shown after subtracting a smooth boxy-elliptical profile model. In (c), we show the color-subtracted image, with results from a lensing model for a singular isothermal ellipse (see text).}
\label{fig:lensmod}
\end{figure}

ESO325-G004 was observed with ACS in January 2005, as part of our program measuring surface-brightness fluctuations in the Shapley Foreground region. The total integration times were 18882 seconds in F814W (22 individual exposures) and 1101 seconds (three exposures) in F475W. The frames were combined using {\sc multidrizzle} in {\sc stsdas} to yield the stacks used in this study (Figure~\ref{fig:image}). Inspection of the images revealed two long, narrow arcs at $\sim$3\,arcsec separation from the galaxy center (Figure~\ref{fig:lensmod}). A third object, less obviously distorted, is present at similar radius. The arcs are prominent even in the much shallower F475W exposure (Figure~\ref{fig:lensmod}a), due to favorable color contrast between the blue arcs and the red light of the foreground elliptical.

For a preliminary photometric analysis of the lens galaxy, we applied {\sc iraf} tasks based on the ellipse-fitting algorithm of Jedrzejewski (1987). The surface photometry model includes the $c_4$ Fourier coefficient, which is necessary to  describe the strong boxiness of the isophotes. All other high-order Fourier terms were forced to zero.  The luminosity profile follows a $R^{1/4}$ law out to at least $\sim$1\,arcmin ($\sim5R_{\rm eff}$). Transformed to the Johnson/Cousins system (following Sirianni et al. 2005), and correcting for extinction and $k$-dimming, the color is $B-I_c\approx2.5$, typical for an old, metal rich stellar population. At the radius of the arcs, the isophotal position angle is 68$^\circ$, and the ellipticity $e=1-\frac{b}{a}\approx0.25$.

Although the arcs are seen clearly in the original images (Figure~\ref{fig:lensmod}a), their visibility is  enhanced by subtracting a model for the lens galaxy light. We have explored a number of approaches to this subtraction. The simplest method uses the ellipse+$c_4$ fits described above, subtracting independent models for the F814W and F475W images. The resulting F814W residual image is shown in Figure~\ref{fig:lensmod}b. This method leaves a strong spoke pattern; there is also a risk of introducing tangential residuals unrelated to lensing. A more robust approach is to use the color contrast between the arcs and the galaxy, subtracting a scaled version of the F814W image from the F475W image. The color residual image (Figure~\ref{fig:lensmod}c) is limited by noise from the shallow F475W exposure, but is systematically very clean; in particular the spoke pattern and many of the point sources are removed.

In Section~\ref{sec:lensmod}, we will show that the observed geometry of arcs `A', `B' and `C' is indeed consistent with a single background source lensed by an elliptical potential.
We have determined approximate magnitudes and colors for the arcs, based on the residual images, as summarized in Table~\ref{tab:arcprops}. In particular, note that the arcs have consistent colors, with $B-I_c=1.10\pm0.05$ (extinction corrected). Such a blue color can be reproduced with star-forming spectral templates at $z\la0.45$, with [O {\sc ii} $\lambda$3727\AA]  contributing to the F475W flux.
The arcs have irregular surface brightness distributions, with numerous `knots', suggestive of star-forming regions. In `B' we can identify at least seven knots, while `A' shows a symmetric structure as expected for a two-image arc, and `C' appears double.

\begin{deluxetable}{cccccc}
\tablecaption{Properties of the arcs}
\tablewidth{0pt}
\tablehead{\colhead{ID}  &
\colhead{radius\tablenotemark{a}}   &
\colhead{length\tablenotemark{a}}   &
\colhead{$F814W$\tablenotemark{b}} &
\colhead{$F475W$\tablenotemark{b}} &
\colhead{$(B-I_c)_{\rm corr}$\tablenotemark{c}}
}
\startdata
A & 2.8 & 3.4 & 22.06 & 22.93 & 1.07 \\
B & 3.4 & 1.5 & 22.92 & 23.88 & 1.21 \\
C & 2.6 & 0.9 & 22.61 & 23.45 & 1.03 \\
\enddata
\tablenotetext{a}{Radius and length (in arcseconds) of an annular segment bounding the $\mu_{\rm F814W}=23$\,mag\,arcsec$^{-1}$ isophote, as used for the photometry.}
\tablenotetext{b}{Integrated magnitude and color, on the Vega system, within the annular segment.}
\tablenotetext{c}{Johnson/Cousins, corrected for galactic extinction.}
\label{tab:arcprops}
\end{deluxetable}

\section{Mass-to-light ratio from lensing}\label{sec:lensmod}

We have used simple mass models to test the assumption that all three images are generated by gravitational lensing, and to derive initial estimates for the mass enclosed within the arcs. To test parametrized forms for the lensing potential, we applied the `curve-fitting' method as implemented within the {\sc gravlens/lensmodel} software (Keeton 2001). This method takes as input a set of points on each observed arc, and optimizes a lensing potential which maps points from each curve to counter-images on the other curves. The input curves were obtained by manually tracing arcs `A', `B' and `C'. To determine a well-defined Einstein radius for an elliptical potential, Rusin, Kochanek \& Keeton (2003), have suggested a scheme based on fitting a singular isothermal sphere (SIS) plus external shear. The model has three parameters, the SIS Einstein radius, and the magnitude and position angle of the shear. This parameter space was explored using a combination of grid-search and simplex optimization methods.

The best SIS+shear fit yields $R_{\rm Ein}\approx2.85$\,arcsec (1.93\,kpc or 0.23\,$R_{\rm eff}$). From this we can determine the mass interior to $R_{\rm Ein}$ to be $M_{\rm Ein}\approx1.4(D_{\rm s}/D_{\rm ls})\times10^{11}\,M_\odot$. Here $D_{\rm s}$ is the angular-diameter distance to the source, and  $D_{\rm ls}$ the angular-diameter distance between lens and source. The enclosed mass measurement thus depends on the source redshift, which is as yet unknown (but see below). 
For a given lens, the source is most likely to be at high redshift, where the surface density of potential sources is very large. In this section, we consider the limiting case of a distant source, such that $D_{\rm s}/D_{\rm ls}\approx1$.
The observed magnitudes interior to $R_{\rm Ein}$ are
$F814W=13.5$ and $F475W=15.6$.
Transforming these to Johnson/Cousins bandpasses following Sirianni et al. (2005), we have
$I_c=12.7$ and $B=15.1$ 
(corrected for extinction and $k$-dimming).
Converting to luminosities we have
$L_{\rm Ein, I_c}=8.0\times10^{10}\,L_{\odot,I_c}$ and
$L_{\rm Ein, B}=3.0\times10^{10}\,L_{\odot,B}$, and the
mass-to-light ratios at the Einstein radius are
$\Upsilon_{\rm Ein, I_c}=1.8\,(M/L)_{\odot,I_c}$ and
$\Upsilon_{\rm Ein, B}=4.7\,(M/L)_{\odot,B}$. Such values are typical for stellar mass-to-light ratios in old populations.
Thus the mass budget within the Einstein radius appears to be dominated by the observed stellar mass\footnote{Given independent information on the stellar population (e.g. from spectroscopic line indices), this result could yield an upper limit on the mass contribution from dark matter associated with \egal\ and/or the surrounding cluster S0740.}. The SIS mass model yields $\sigma_{\rm SIS}=310$\,\kms, in excellent agreement with the measured stellar velocity dispersion.

Considering the morphology of the galaxy, and the expectation that the mass within $R_{\rm Ein}$ will be dominated by stars, a more realistic model for the lensing potential is the Singular Isothermal Ellipse (SIE; e.g. Korman, Schneider \& Bartelmann 1994).
In fitting the SIE model, we force the center of the potential to align with the observed galaxy, but allow the mass normalization, position angle and ellipticity to vary. Formally, the best fitting model has ellipticity $e=0.28$ with position angle 68$^\circ$, very similar to the equivalent parameters for the galaxy light. Figure~\ref{fig:lensmod}c shows the results of this best-fitting SIE model. The reconstructed source position is $\sim$0.3\,arcsec north of the lens center (small circle), and likely straddles the inner `astroid' caustic of the lens model (inner cusped curve). The figure shows the predicted image locations for such a source. Qualitatively, the SIE potential reproduces the broad features of the observed geometry, and confirms image `C' as a counter-image of `A' and `B'. The long arc `A' comprises two images crossing the critical curve (outer ellipse). 
While broadly successful, it is also clear that the model does not match the data perfectly at a more detailed level.
In particular, the SIE does not match the relative lengths of the three observed arcs.  For `C', moreover, the model predicts a position displaced towards `B', relative to the observed location. To improve the reconstruction, more sophisticated models could take into account the intrinsic morphology of the source, the boxiness of the lens galaxy, and a possible external shear term. 

\section{Discussion}

It is interesting to consider whether the lensing configuration observed in \egal\ is intrinsically unusual, or whether many other nearby galaxies might reveal such bright arcs when observed in sufficient detail.
The cross-section for four-image lensing under the SIE mass-model is $\sim$1\,arcsec$^2$, i.e. roughly the area of the `astroid' caustic on the source plane. The estimated magnification factors for the separated arcs `B' and `C' are $\sim$4, suggesting the unlensed magnitude of the source is $I_c\sim23.75$. To this limit, the integrated I-band galaxy counts are $\sim10^5$\,deg$^{-2}$ (Postman et al. 1998), so the probability that a given galaxy has such a bright background source aligned for quadruple imaging is $\sim$0.6\%.
To estimate the total number of massive galaxies available to serve as low-redshift lenses, we use the 2MASS J-band luminosity function of Cole et al. (2001). \egal\ has $M_J-5\log h=-24.16$, and thus a luminosity $\sim7L_J^\star$. Integrating the luminosity function, parametrized as a Schechter function, the space density of galaxies above $7L_J^\star$ is $2.8\times10^{-5}\,{\rm Mpc}^{-3}$ (for $h=0.71$). 
This density implies $\sim10^4$ galaxies, at least as luminous as \egal, within $z=0.1$. Combining these estimates, we find the total expected number of `similar' low-redshift lens systems to be $\sim$60, with only $\sim$3 at the distance of \egal\ or closer.

As noted above, the lens model normalization depends on the source redshift, through the factor $D_{\rm s}/D_{\rm ls}$. We have re-examined the raw spectra obtained by Smith et al. (2000) at the Anglo-Australian Telescope, which cover the range $4925-5740$\,\AA, with slit intercepting Arc `C'. No anomalous emission lines are seen at the expected position of the arc, which at face value suggests $z_{\rm src}>0.54$ and $z_{\rm src}<0.32$ (absence of [O {\sc ii} $\lambda$3727\AA]), but also 
$z_{\rm src}>0.15$ (absence of [O {\sc iii} $\lambda$5007\AA] and H\,$\beta$). Over the redshift interval $0.15-0.32$, the lensing mass correction factor $D_{\rm s}/D_{\rm ls}$ ranges from 1.30 to 1.13. It is however quite possible that the exposures were too short (600\,sec), and the slit too wide (3\,arcsec), to detect emission from the arc against the high background of the lens galaxy.

Finally, we note that our images show four extremely faint tangential arc candidates at greater separation from \egal\ (radius 9.4\,arcsec, at position angles $-95^\circ$, $90^\circ$, $-35^\circ$, $150^\circ$). The surface brightness of these features is very low ($\ga24.5$\,mag\,arcsec$^{-2}$ in I). If confirmed, the outer arcs could provide additional constraints on the mass distribution at larger radius, and on the relative contribution of the group potential to the lensing mass.

\section{Conclusions}\label{sec:concs}

We have discovered strong gravitational lensing by the elliptical galaxy \egal, which at $z=0.0345$ is the nearest known galaxy-scale lens. The multiply-imaged background source appears to be a star-forming galaxy, with prominent substructure. If the source is very distant, the arc radius is consistent with a stellar-dominated mass-to-light ratio within radius $\sim{}\frac{1}{4}R_{\rm eff}$. An elliptical isothermal mass model recovers the position angle and ellipticity of the galaxy, independent of the observed luminosity. The best-fit mass scale is consistent with the measured stellar velocity dispersion. Future modeling of the system should incorporate structural information from the clumpy arc morphology. We intend to obtain integral-field spectroscopy for \egal, with 8m-class telescopes, to secure the source redshift and enclosed mass estimate. Additionally, these observations will yield extended stellar dynamics for the lens galaxy, and enable high-contrast imaging of the arcs by emission-line mapping.

\acknowledgments

RJS thanks  Mike Hudson and Laura Parker for useful discussions about this work, and the
Anglo-Australian Observatory for retrieving the raw AAT spectra of \egal.
This research has made use of the NASA/IPAC Extragalactic Database (NED)
which is operated by the Jet Propulsion Laboratory, California Institute of
Technology, under contract with the National Aeronautics and Space Administration.
This publication makes use of data products from the Two Micron All Sky Survey, which is a joint project of the University of Massachusetts and the Infrared Processing and Analysis Center/California Institute of Technology, funded by the National Aeronautics and Space Administration and the National Science Foundation.

\end{document}